\documentstyle[12pt]{article}
\begin{document}
\renewcommand{\theequation}{\thesection.\arabic{equation}}
\begin{titlepage}
\renewcommand{\thefootnote}{\fnsymbol{footnote}}
\begin{flushright}
\large{FAU-TP3-95/3} \\
\large{April 1995}
\\
\end{flushright}
\vspace*{3.5cm}
\begin{center} \LARGE
{\bf  Higgs~Mechanism~and~Symmetry~Breaking without Redundant Variables}
\end{center}
\vspace*{1.0cm}
\begin{center}

{\bf D. Stoll$\,^1$ and M. Thies$\,^2$}

\vspace*{0.2cm}

{\em Institute for Theoretical Physics, University of
Erlangen-N\"{u}rnberg,
\\
Staudtstr. 7, 91058 Erlangen, Germany}
\vspace*{0.2cm}

\end{center}

\vspace*{2.0cm}

\begin{abstract}

The Higgs mechanism is reconsidered in the canonical Weyl gauge
formulation of quantized gauge theories, using an approach in which
redundant degrees of
freedom are eliminated. As a consequence, its symmetry aspects
appear in a different light. All the established physics
consequences of the Higgs mechanism are recovered without
invoking gauge symmetry breaking.
The occurence of massless vector bosons in non-abelian Higgs models
is interpreted as
signal of spontaneous breakdown of certain global symmetries.
Characteristic differences between the relevant
``displacement symmetries'' of QED and the Georgi Glashow model
are exhibited. Implications for the symmetry aspects of the
electroweak sector of the
standard model and the interpretation of the physical photon as Goldstone
boson are pointed out.

\end{abstract}

\vskip 4.0cm

$^1$ stoll@theorie3.physik.uni-erlangen.de

$^2$ thies@theorie3.physik.uni-erlangen.de

\end{titlepage}

\setcounter{page}{2}

\addtocounter{footnote}{0}

\newpage

\setcounter{equation}{0}
\section{Introduction}
Gauge field theories have established themselves as the key concept in
formulating
and understanding all fundamental interactions. They have been
successfully applied to the description of perturbative processes,
both in the abelian theory (QED) without selfcoupling of gauge
degrees of freedom and in non-abelian theories like QCD, where
the coupling of gluons amongst each other becomes relevant. The most
spectacular success of gauge theories was however the prediction and
subsequent experimental confirmation of
massive vector bosons in the
Glashow-Weinberg-Salam (GWS) model \cite{GWS,Arnison}, made possible by
the discovery of the
Higgs mechanism \cite{Englert}--\cite{Kibble}. The fact that gauge bosons
acquire
mass is rightfully considered to be an extremely important phenomenon.
It has attracted a lot of attention,
also in connection with attempts
to further unify the fundamental electromagnetic, weak and strong
interactions. By way of contrast, the fact that gauge theories are capable
of generating massless vector particles has usually been taken for granted
and not considered worth discussing. As a result,
somewhat surprisingly, there does not seem to be a consensus in the
literature on the fundamental question why the observed photon is
massless.

In most of the literature including the standard textbooks, one considers
the absence of a mass term in the Lagrangian or Hamiltonian as indicative of
whether massless vector bosons will appear in the spectrum.
The corresponding gauge group
(or subgroup) is then said to be ``unbroken''. However, it is well known
since Schwinger \cite{Schwinger}
that mass can appear dynamically, without any conflict with gauge
invariance. Moreover, such a scheme gives no clue why
only unbroken {\em abelian} gauge groups seem to
give rise to massless photons. Yang Mills theory or QCD, which are also
considered as
``unbroken'', do not exhibit massless vector particles, a fact which then
has to be attributed to the poorly understood confinement phenomenon.

A very different line of reasoning goes back to Guralnik
\cite{Guralnik2} and has been periodically revived since then
(cf. \cite{LNOT12} and references therein).
In these works,
the idea is put forward
that the photon of QED can be viewed as Goldstone
boson. The symmetry which is spontaneously broken is related to
gauge symmetry -- it is that
part which is not used up in eliminating redundant degrees
of freedom, the invariance under ``large'' gauge transformations.
These considerations have the appealing feature that only one single
mechanism, the breaking of global symmetries
\footnote{``Global symmetry'' refers to transformations which can be defined
in terms of a finite number of constant parameters and should not be
confused with the so-called ``gauge transformations of the first kind''.},
is responsible for the existence of massless states, in ungauged as well as
in gauged theories. Nevertheless, such ideas have never found wide
acceptance,
presumably because there seems to be no useful order parameter
associated with the symmetry breakdown. Moreover, this mechanism has
never been confronted with the fact that also in non-abelian gauge
theories with simple gauge groups like the Georgi Glashow model
\cite{Georgi}, massless vector mesons can appear.

Recently, the issue of photons as Goldstone bosons has again been taken up
in the context of QED, both in a
dual, ``magnetic'' formulation \cite{Kovner} and in the conventional,
``electric'' one \cite{LNOT12}. Extending the work of Ref. \cite{LNOT12},
we propose to study non-abelian Higgs models in a canonical framework
where all (redundant) gauge degrees of
freedom are eliminated. This then should enable us to identify
residual global symmetries which \ -- by spontaneous breakdown -- \ can
account for the observed massless particles.
For this purpose we follow the approach to
first quantize the Hamiltonian in the Weyl gauge ($A_0=0$) on a
3-dimensional torus and to eliminate afterwards
all gauge variant degrees of freedom by means of ``unitary
gauge fixing transformations''\cite{LNOT12}.
Use of a torus (i.e., periodic boundary conditions in a box) has proven
helpful both to control infrared divergences and to clearly separate
the ``large'' gauge transformations from the ``small'' ones,
generated by the Gauss law operator. One can then use homotopy considerations
to classify mappings of the torus $T^3$ into the U(1) gauge group.
(Such topological considerations are more familiar from non-abelian gauge
theories, in particular QCD, where they are crucial for understanding
the $\theta$ vacuum angle and instantons in the canonical framework
\cite{Jackiw}.)
By generalizing the methods of Ref. \cite{LNOT12}, it proves possible
to study systematically the most important Higgs models,
irrespective of the gauge groups or the representation
for the Higgs fields. Technically, the basis for such a unified description
is the use of variables taking values in the gauge group.
The resulting methods to eliminate gauge variant variables
from the Hamiltonian were already shown to reduce the complexity of
performing unitary transformations in Yang-Mills
theories
\cite{Stoll}. They will turn out to be efficient also in the case of Higgs
models where they can be applied rather easily.

We should like to point out another conceptual difference between
this approach and the standard one using redundant variables, which
is relevant for the interpretation of the Higgs mechanism.
As a consequence of the reduction of (``small'') local gauge transformations
to unity
in the physical Hilbert space after eliminating gauge variant degrees of
freedom, only global symmetries can survive.
The notion of spontaneous breakdown of local gauge symmetry
is therefore no issue in our work. It was in fact pointed out by Elitzur
\cite{Elitzur}
some time ago that a local gauge symmetry cannot
be spontaneously broken. The reason is the same as
the one which forbids spontaneous symmetry breaking in quantum mechanics
as opposed to quantum field theory:
Only symmetries which involve infinitely many degrees of freedom
can be spontaneously broken.
Local gauge symmetries can act on a finite number of degrees of freedom,
namely those which are available at one point in space, and therefore
do not satisfy this criterion.
Indeed, non-perturbative investigations of the Higgs mechanism
in the temporal gauge have shown that it is not accompanied by any
symmetry violating order parameter
\cite{Froehlich}. There is an ongoing debate about the interpretation of the
Higgs mechanism in terms of gauge symmetry breaking, as
can also be read off from titles of publications such
as ``Gauge-invariant signal for gauge-symmetry breaking'' \cite{Dolan}
or ``Spontaneously unbroken symmetry and gauge-invariant
effective action'' \cite{Lawrie}. This discussion has to do with
the interpretation of the Higgs mechanism, rather than with its
practical consequences everybody agrees upon. We will
try to contribute to this discussion
by re-examining the abelian Higgs mechanism first. In that case, it is
easy to understand why seemingly different conceptual approaches
at the end lead to the same observable effects.

A remark about a limitation of our approach is in order:
Throughout this paper,
we will not consider questions of UV-regularization or
renormalization, but instead deal with the formulae in a rather
symbolic manner.
Our tacit assumption is that the
symmetry aspects we are
studying are not particularly sensitive to the UV behaviour
of the theories considered. Although such an
assumption is presumably better justified in
the Higgs phase of gauge theories than in other phases, we cannot
a priori rule out that certain conclusions might be altered
in a more rigorous treatment.

Finally, we recall that the guiding principle followed
here \ -- to eliminate the redundant gauge degrees of freedom --\  is
not the only possibility.
In dealing with gauge theories, many workers prefer to
even increase the number of variables further and then try to understand the
symmetries in the context of BRST quantization \cite{BRST}. Although
it would be very
interesting to establish the connection between these two opposite
attitudes,
this has not yet been done, and we follow the first route in the
present work.

This paper is organized as follows:
In Sect. 2, we briefly review the situation for the abelian
Higgs model within the canonical approach and contemplate its symmetry
aspects. We
compare our derivation to the conventional one, trying
to pinpoint why the latter yields the same results in spite of
conceptual differences.
In Sect. 3, we develop the unitary/Coulomb gauge representation
of the Georgi Glashow model, a formulation exclusively in terms of physical
variables. Sect. 4 contains a digression on gauge invariant
operators like the 't Hooft tensor \cite{Hooft74}.
This preparation is necessary in order to understand firmly the residual
displacement symmetry of the Georgi Glashow model in Sect. 5
and contrast it with the QED case.
In Sect. 6, we carry out the quantum mechanical gauge fixing for
Higgs models with fundamental scalars, considering first a pure SU(2) model
and then a U(1)$\times$SU(2) model, the bosonic sector of the GWS theory.
We analyse the residual global symmetries in both cases.
Finally, Sect. 7 is devoted to a summary and our conclusions.

\setcounter{equation}{0}
\section{Reminder of the abelian Higgs mechanism}

Since we propose to
extend the methods developed in Ref. \cite{LNOT12} to non-abelian
Higgs models, it may be
worthwhile to recall how one eliminates redundant variables in
the simpler case of the abelian model first.
This should render the present work essentially self-contained.
We will also point out the difference to the
conventional
treatment of the Higgs mechanism, not in the results (which are
indistinguishable), but concerning the interpretation of gauge symmetry
aspects.

In the canonical Weyl gauge formulation ($A^0=0$), a complex scalar
field coupled to a
U(1) gauge field is described by pairs of conjugate
variables ($\varphi_i,\pi_i$)
and ($\vec{\cal A},-\vec{\cal E}$).
Periodic boundary conditions are imposed
on all fields, i.e., 3-space is compactified to a torus.
Throughout this paper, we suppress the $\vec{x}$ arguments
of fields and other operators whenever possible.
The Hamiltonian density
\begin{equation}
{\cal H} = \frac{1}{2} (\vec{\cal E}^{\,2} + \vec{\cal B}^{\,2}
) + \pi^{\dagger} \pi + (\vec{D} \varphi )^{\dagger} (\vec{D}\varphi)
+ V(\varphi^{\dagger} \varphi)
\label{2.1}
\end{equation}
involves the standard magnetic field $\vec{\cal B} = \vec{\nabla}\times
\vec{\cal A}$ and covariant derivative $\vec{D} = \vec{\nabla}
-ie\vec{\cal A}$. The Hamiltonian is invariant under time independent
local gauge transformations characterized by a function $\beta(\vec{x})$,
\begin{equation}
\vec{\cal A}\to \vec{\cal A} + \vec{\nabla} \beta \ , \qquad
\varphi \to e^{ie\beta} \varphi \ .
\label{2.1a}
\end{equation}
As for the complex scalar field, the following definitions
are useful,
\begin{equation}
\varphi = \frac{1}{\sqrt{2}}(\varphi_1 + i \varphi_2)
= \frac{1}{\sqrt{2}} \chi \hat{\varphi} = \frac{1}{\sqrt{2}}
\chi e^{ie\Delta} \ .
\label{2.2}
\end{equation}
Here,
$\chi$ is a hermitian, ``radial'' field ($\chi^2=\varphi_1^2+\varphi_2^2$)
singled out by the fact that it is gauge invariant. (To introduce so
many different definitions at once may seem uneconomical, but will pay
off very soon. Moreover, similar notations will be used
again in the non-abelian Higgs models.)
Gauss's law is imposed as a constraint on the physical states,
\begin{equation}
(-\vec{\nabla}\vec{\cal E} + e \rho )| \ \rangle = 0 \ ,
\label{2.3}
\end{equation}
where the charge density of the matter field,
\begin{equation}
e\rho = ie(\varphi^{\dagger} \pi - \pi^{\dagger}\varphi  ) \ ,
\label{2.4}
\end{equation}
generates local gauge transformations of $\varphi$.
It is convenient to introduce the (dynamical) set of orthonormal
vectors
\begin{equation}
v^1 = \left( \begin{array}{rr} -\hat{\varphi}_2 \\ \hat{\varphi}_1
\end{array} \right) \ , \qquad
v^2 = \left( \begin{array}{rr} \hat{\varphi}_1 \\ \hat{\varphi}_2
\end{array} \right) \ ,
\label{2.5}
\end{equation}
in terms of which
\begin{equation}
\rho = - \chi \sum_{i=1,2}v^1_i \pi_i := - \chi (v^1,\pi) \ .
\label{2.5a}
\end{equation}
Provided that $\chi \neq 0$,
the Gauss law can trivially be resolved,
\begin{equation}
(v^1,\pi)| \ \rangle = - \frac{1}{e\chi} \vec{\nabla}\vec{\cal E}| \
\rangle \ .
\label{2.6}
\end{equation}
The other component of $\pi$, $(v^2,\pi)$, is the ``radial'' momentum
unconstrained by Gauss's law and will be denoted by $p$.
In the physical sector, the Higgs field
kinetic energy density can then be replaced by
\begin{equation}
\langle \ | \pi^{\dagger} \pi | \ \rangle = \langle \ |
\left( \frac{1}{2} p^{\dagger}p + \frac{1}{2(e\chi)^2} (\vec{\nabla}
\vec{\cal E})^2 \right) | \ \rangle \ .
\label{2.7}
\end{equation}
Note that $p$ is a non-hermitian operator, reflecting the presence of
a non-trivial Jacobian when going from cartesian to (plane) polar coordinates.
As discussed in \cite{LNT,Shifman}, such Jacobians
and the corresponding boundary conditions on ``radial wave functionals''
have rather drastic consequences on the dynamics in the case of QCD.
Here, this aspect will
not play any comparable role. In the Higgs phase, we will assume
that fluctuations of the variable $\chi$ about its classical value
are too small to lead into the vicinity of $\chi = 0$. Otherwise,
our way of resolving Gauss's law is problematic anyway.

The next step consists in transforming away the variables conjugate
to $(v^1,\pi)$, i.e., the
phase of the Higgs field. This is achieved by transforming the
Hamiltonian (\ref{2.1}) via the unitary gauge fixing transformation (UGFT)
\begin{equation}
{\cal U}_{\Delta} = \exp \left(-i\int d^3 x \vec{\cal E} \vec{\nabla}
\Delta \right) \ ,
\label{2.8}
\end{equation}
with $\Delta$ the phase of the Higgs field as defined in eq. (\ref{2.2}).
After projection onto the physical sector (cf. (\ref{2.7})), only one term
in $\langle \ |\cal{H}|\ \rangle$,
$(\vec{D} \varphi)^{\dagger} (\vec{D}\varphi)$, is affected by
the unitary transformation. As a result,
$\varphi$ simply gets replaced by $\chi/\sqrt{2}$, as if we had
``fixed the gauge'' classically. The Hamiltonian density
$\tilde{\cal H} = {\cal U}_{\Delta}{\cal H}\,{\cal U}_{\Delta}^{\dagger}$
in the ``unitary gauge representation'', projected onto the physical
states, becomes
\begin{equation}
\tilde{\cal H} = \frac{1}{2} (\vec{\cal E}^{\,2} + \vec{\cal B}^{\,2}
) + \frac{1}{2}p^{\dagger}p + \frac{1}{2(e\chi)^2} (\vec{\nabla}\vec{\cal E}
)^2 + \frac{1}{2}(\vec{\nabla}\chi)^2 +
\frac{1}{2} (e\chi)^2\vec{\cal A}^{\,2}
+ V(\chi^2/2) \ .
\label{2.9}
\end{equation}
The gauge has not been fixed beyond the Weyl gauge. All we have done
is to resolve Gauss's law and perform a unitary transformation
which eliminates the angular variables of the Higgs field.
If $\chi$ assumes a non-vanishing expectation value, we recover
the standard mass term for the vector field in (\ref{2.9})
with mass $m=e\langle\chi\rangle$.

Let us now discuss the symmetry aspects of the abelian Higgs mechanism
and compare our approach with the
conventional derivation. The above procedure has left no apparent residual
gauge symmetry.
This fact may mean one out of two things:
Either, all gauge transformations
are reduced to {\bf 1} and everything is trivially gauge invariant, or
the coordinates have been chosen implicitly in a way which already has built
in the breaking of certain gauge symmetries, so that the corresponding
symmetry is hidden.

In order to understand in which situation we are, it is preferable
to switch
over to a less ``biased'' gauge, the Coulomb gauge (the unitary gauge
is extremely singular in the phase where there are massless photons).
Classically, the Coulomb gauge condition $\vec{\nabla}
\vec{A}= 0$ does not fix the gauge completely, but leaves the freedom
of transformations involving
either constant or linearly $\vec{x}$-dependent gauge functions,
\begin{equation}
           \vec{A} \to \vec{A}+\frac{2\pi}{eL}\vec{n} \ , \qquad
\varphi \to  \exp \left\{ie\left(\beta_0 + \frac{2\pi}{eL} \vec{x}\vec{n}
\right) \right\}\varphi \ .
\label{2.10}
\end{equation}
Here, $L$ is the extention of the torus and the $n_i$ are integers.
The existence of these ``Gribov copies'' also manifests itself
in the framework of quantum mechanical gauge fixing \cite{LNOT12}. There, one
finds that the Hamiltonian in the Coulomb gauge representation
commutes with the unitary operators
\begin{equation}
\Omega[\vec{n},\beta_0] = \exp \left\{ -i \left( \frac{2\pi}{eL}
\vec{\cal D}\vec{n}
+ e Q \beta_0 \right) \right\} \ .
\label{2.11}
\end{equation}
The total charge has been introduced as
$eQ=e\int d^3x \rho$, whereas $\vec{\cal D}$ denotes the
gauge invariant operator
\begin{equation}
\vec{\cal D} = \int d^3 x \left( \vec{E} + e \vec{x} \rho \right)  \ .
\label{2.12}
\end{equation}
In analogy to the displacement vector in macroscopic, classical
electrodynamics,
$\vec{\cal D}$ has been called
``displacement vector" in Ref. \cite{LNOT12}, and the corresponding
residual gauge symmetry
``displacement symmetry".

The displacement symmetry fits nicely into the general theoretical
expectation according to which
only topologically non-trivial, so-called ``large" gauge
transformations should survive in the
physical sector, since the ``small'' ones can be generated by the Gauss law
operator \cite{Jackiw}. Indeed, the
integer displacements $\vec{n}$ in eq. (\ref{2.11}) are nothing
but the winding numbers of the mapping
$T^3 \to $U(1) for each
generating circle of the torus $T^3$.
In the
limit $L \to \infty$, the transformation (\ref{2.11}) becomes
a continuous symmetry which is spontaneously broken in standard QED;
photons are the corresponding
Goldstone bosons. In the Higgs phase,
there is evidence (although no rigorous formal proof) that
the displacement symmetry is realized in the Wigner Weyl mode \cite{LNOT12}.
  From this point of view,
the fact that the vector meson acquires a mass seems very natural.

The global gauge transformations generated by $Q$ in eq. (\ref{2.11})
raise a more delicate issue. So far,
it has been tacitly assumed that
together with the local Gauss law, also the global one (i.e., the Gauss law
integrated over all space)
is part of the definition of the theory and should be imposed strictly
on the physical states. In the Coulomb gauge representation, there appears
a residual Gauss law, the neutrality condition
\begin{equation}
eQ|\ \rangle = 0 \ .
\label{2.13}
\end{equation}
It is interesting to ask whether this condition can be relaxed, or
perhaps even has to be abandoned
in the phase where $\langle \chi
\rangle \neq 0$.
One observation has to be kept in mind, though:
As discussed in \cite{LNOT12}, one cannot have simultaneously
translational invariance and invariance under displacements, unless
one restricts oneself to the neutral sector.
Denoting the generators of the
relevant symmetries by $Q$ (gauge transformations of the first kind),
$\vec{\cal D}$ (displacements) and $\vec{P}$ (translations),
this follows at
once from the algebraic relation
\begin{equation}
[{\cal D}_i ,P_j ] = ie Q \delta_{ij} \ .
\label{2.14a}
\end{equation}
Physically, it
reflects the well-known fact that only for neutral systems the electric
dipole moment is translationally invariant.

One is left then with three
options in the Higgs phase: One can break $Q$ together with $\vec{\cal D}$,
$Q$ together with $\vec{P}$, or not
break any symmetry at all. The first two possibilities lead to
a number of unwanted Goldstone bosons and are hard to reconcile
with the findings in non-relativistic many-body systems or the
Schwinger model \cite{LNOT12}. Since we anyway would have difficulties
to understand the physics meaning of a conservation law which holds
locally, but not globally, we favour the third option where
the Higgs phase is characterized by a perfect symmetry.
This is the simplest interpretation in a framework which uses only
unconstrained variables.

Let us now briefly reconsider the derivation of the Higgs mechanism
as found in many textbooks. One typically
starts the discussion from the ungauged scalar field theory with
spontaneous breakdown of the global U(1) symmetry and non-zero
vacuum expectation value $\langle \varphi \rangle$. When the coupling
to the gauge field is turned on, it is assumed that
the symmetry breakdown is not affected. Thus, one
continues to work with the same expectation value $\langle \varphi \rangle$.
The widespread opinion that the Higgs mechanism implies breakdown
of local gauge symmetry has its origin in such a reasoning.
Historically, this line of thought was very important for the
discovery of the Higgs mechanism, starting out from spontaneously
broken non-gauge theories. It has led to a wealth of correct physics
results. How can one understand the fact that it yields the same
results as the above approach in terms of unconstrained variables,
which is conceptually quite different and does not depend on
gauge symmetry breakdown? Comparing the two approaches, we recognize
that the essence of our method consists in first
fixing the gauge, then shifting the scalar field
(where gauge fixing means of course the transition to gauge invariant
variables),
\begin{equation}
\varphi \ \ \stackrel{\rm fix}{\longrightarrow} \ \ \varphi'
\ \ \stackrel{\rm shift}{\longrightarrow} \ \
\langle \varphi' \rangle +
\tilde{\varphi}' \ .
\label{2.18}
\end{equation}
In the unitary gauge, $\varphi'=\chi$, and no symmetry is broken by
the non-vanishing expectation value. Clearly, the scheme (\ref{2.18})
is not specific for
the canonical approach. If we apply it to the Lagrangian approach
at the classical level, we get the standard results for the Higgs
mechanism without any effort and without invoking (local) gauge symmetry
breaking (this has occasionally been done in the literature, see
e.g. \cite{Lawrie,Coleman}).
The conventional approach differs from this one by the interchange
of the two steps -- one shifts the field before
gauge fixing,
\begin{equation}
\varphi \ \ \stackrel{\rm shift}{\longrightarrow} \ \ \langle \varphi
\rangle + \tilde{\varphi} \ \ \stackrel{\rm fix}{\longrightarrow}
\ \ \langle \varphi \rangle  + \tilde{\varphi}' \ .
\label{2.19}
\end{equation}
Here, one is led to the notion that the local gauge
symmetry is broken. Nevertheless, comparing the results of
the operations (\ref{2.18}) and (\ref{2.19}), one still can
get the same final answer, provided one takes an expectation
value $\langle \varphi \rangle$ consistent with the chosen
gauge. How can this happen if one
introduces $\langle \varphi \rangle$ {\em before}
committing oneself to a specific gauge? In practice, the Higgs
mechanism is
most conveniently discussed in the unitary gauge. In this gauge,
the expectation value of $\varphi$ has exactly the same form
as in the ungauged Higgs model. A moment's thought shows that this
holds true in the non-abelian case as well, for scalar fields
in the fundamental or adjoint representation.
Hence, if one simply
takes over $\langle \varphi
\rangle $ from the ungauged Higgs model and later on uses the unitary
gauge, one gets the same result as if one had performed
the two steps indicated in (\ref{2.19}) in the reverse order.

Thus, we confirm previous findings that it is not necessary to invoke
breaking of local gauge invariance in order to get all the physics
consequences of the Higgs mechanism. In addition, we have given a
simple explanation why the conventional derivation of the Higgs
mechanism yields the correct result.
In a formulation without redundant variables, we have no other choice
than to proceed according to the scheme (\ref{2.18}). In the remainder of
this paper, this strategy will be applied to non-abelian
Higgs models.

\setcounter{equation}{0}
\section{Georgi Glashow model in the unitary/Coulomb gauge representation}

Consider the Georgi Glashow model \cite{Georgi} which consists of
self-interacting, scalar matter fields $\phi^{a}$ in the adjoint
representation of SU(2), minimally
coupled to SU(2) Yang-Mills fields $\vec{A}^{\,a}$.
The momenta conjugate to $\phi^{a}$ and $\vec{A}^{a}$ appearing in the
canonical Weyl gauge formulation will be denoted by
$\pi^{a}$ and $-\vec{E}^{a}$, respectively.
Starting point is
the Hamiltonian density
\begin{equation}
{\cal H}  =  \frac{1}{2} \pi^{a}\pi^{a}  + V( \phi^{a} \phi^{a})
 + \frac{1}{2} ( \vec{D} \phi )^{a} (\vec{D}\phi )^{a}
+ \frac{1}{2} ( \vec{E}^{a} \vec{E}^{a} + \vec{B}^{a} \vec{B}^{a} ) \ ,
\label{3.1}
\end{equation}
with the non-abelian magnetic field and the covariant derivative given by
\begin{eqnarray}
\vec{B}^{a} & = & \vec{\nabla} \times \vec{A}^{a} + \frac{1}{2} g
\epsilon^{abc}
\vec{A}^{\,b} \times \vec{A}^{c} \ , \nonumber \\
(\vec{D} \phi )^{a} & = & \vec{\nabla} \phi^{a} + g \epsilon^{abc}
\vec{A}^{\,b} \phi^{c} \ .
\label{3.2}
\end{eqnarray}
Gauss's law will be imposed as a constraint on the physical states,
\begin{equation}
G^{a}| \ \rangle = \left( G_{\rm rad}^{a} + g \rho_{\rm matt}^{a}
\right) | \ \rangle = 0  \ .
\label{3.3}
\end{equation}
Here, the radiation Gauss law operators and the SU(2) charge
densities are
\begin{equation}
G_{\rm rad}^{a}  =  - \vec{\nabla} \vec{E}^{a}  + g \rho_{\rm rad}^a
\ , \qquad \ \ \rho_{\rm rad}^a =
 - g \epsilon^{abc}
\vec{A}^{\,b} \vec{E}^{c}  \ , \qquad \ \
 \rho_{\rm matt}^{a}  =   \epsilon^{abc} \phi^{b} \pi^{c}  \ ,
\label{3.4}
\end{equation}
and we recall the basic commutation relations
\begin{equation}
\left[ G_{\rm rad}^{a}(\vec{x}), G_{\rm rad}^{b}(\vec{y}) \right] =
i \epsilon^{abc} G_{\rm rad}^{c}(\vec{x}) \delta^{(3)}(\vec{x}-\vec{y}) \ .
\label{3.4a}
\end{equation}
We shall derive the unitary gauge representation,
following the general method developed in Ref. \cite{LNOT12}
and applied to a non-abelian gauge theory (QCD with fundamental
fermions in an axial gauge) in \cite{LNT}.
We choose to eliminate $\pi $, the momentum conjugate to the matter field,
to the extent allowed by the structure of Gauss's law.
In order to resolve the Gauss law, we
diagonalize the matrix multiplying $\pi$ in $\rho_{\rm matt}$,
\begin{equation}
g \epsilon^{abc} \phi^{b} v_{n}^{c} = i \mu_{n} v_{n}^{a}  \qquad
(n=1,2,3) \ .
\label{3.5}
\end{equation}
This problem can easily be solved with standard vector algebra, as is
seen by rewriting it in vector form in internal space.
We use the unit vectors $\vec{e}_r,\vec{e}_{\vartheta},
\vec{e}_{\varphi}$ familiar from polar coordinates with the property
\begin{equation}
\vec{e}_r \times \vec{e}_{\vartheta} = \vec{e}_{\varphi} \qquad
(+ \mbox{cyclic}) \ .
\label{3.6}
\end{equation}
Identifying the direction of the Higgs field with $\vec{e}_r$
and denoting its length by $\chi$,
eq. (\ref{3.5}) is rewritten as
\begin{equation}
g \chi \vec{e}_r \times \vec{v}_n = i \mu_n \vec{v}_n
\label{3.7}
\end{equation}
with the solutions
\begin{eqnarray}
\vec{v}_{1,2} & = & \frac{1}{\sqrt{2}} (\vec{e}_{\vartheta} \pm i
\vec{e}_{\varphi}) \ , \qquad \mu_{1,2} = \mp g \chi
\nonumber \\
\vec{v}_3 & = & \vec{e}_r \ , \qquad \qquad \qquad \ \ \ \ \ \mu_3=0 \ .
\label{3.8}
\end{eqnarray}
We expand $\pi$ in the basis of the $\vec{v}_n$ and substitute it back
into the Gauss law,
\begin{equation}
\left( G^a_{\rm rad} + \sum_n i \mu_n v_n^a (v_n, \pi) \right)
|\ \rangle = 0 \ .
\label{3.9}
\end{equation}
Projecting this equation onto $\vec{v}_{1,2}$ from the left allows
us to express
the 1,2 components of $\pi$ in the new dynamical basis directly in terms of
$G_{\rm rad}$. Since $\mu_3=0$, the 3rd component $(v_3,\pi)$ is not
constrained by the Gauss law and survives as physical operator.
It corresponds to the radial momentum operator and
will again be denoted by $p$. Thus, in the physical
space, the matter field kinetic energy is equivalent to
\begin{equation}
\langle \  | \frac{1}{2} \pi^a \pi^a |\  \rangle
=\langle \ | \left( \frac{1}{2} p^{\dagger}p
+ \frac{1}{2(g\chi)^2} \sum_{a=1,2} G^a_{\rm rad}G^a_{\rm rad} \right)
|\   \rangle \ .
\label{3.10}
\end{equation}
The projection of the Gauss law onto $v_3$ yields the residual
constraint
\begin{equation}
(v_3,G_{\rm rad}) |\ \rangle = \hat{\phi}^a G^a_{\rm rad}
|\ \rangle = 0 \ .
\label{3.11}
\end{equation}

In the next step,
we have to eliminate the angular variables of the Higgs field by
a unitary transformation. Introduce the SU(2) matrix $e^{ig\Delta}$ via
\begin{equation}
\phi = \chi \hat{\phi} = \chi e^{ig\Delta} \frac{\sigma^3}{2} e^{-ig\Delta}
\ .
\label{3.12}
\end{equation}
Then, the gauge
fixing transformation
appropriate for the unitary gauge is
\begin{equation}
{\cal U}_{\Delta} = \exp \left\{ -i \int d^{3}x {\cal G}_{\rm rad}^a
\Delta^{a}  \right\}  \ ,
\label{3.13}
\end{equation}
with the definition and local commutation
relations of ${\cal G}_{\rm rad}^a$,
\begin{eqnarray}
{\cal G}_{\rm rad}^a &=& \vec{E}^{a} \vec{\nabla} + g \rho_{\rm rad}^a
\ , \nonumber \\
\left[ {\cal G}_{\rm rad}^a(\vec{x}) , {\cal G}_{\rm rad}^b(\vec{y}) \right]
&=& i\epsilon^{abc}{\cal G}_{\rm rad}^c(\vec{x})
\delta^{(3)}(\vec{x}-\vec{y})  \ .
\label{3.13a}
\end{eqnarray}
${\cal U}_{\Delta}$ is a gauge transformation on all the
radiation variables
with gauge function
$e^{ig \Delta}$,
does not affect $p$ and eliminates the angular variables
of the matter field from the Hamiltonian, leaving only
the radial variable $\chi$. The result for the Hamiltonian density
$\tilde{\cal H} = {\cal U}_{\Delta} {\cal H}\, {\cal U}_{\Delta}^{\dagger}$
in the physical sector is
\begin{eqnarray}
{\cal H} & = & \frac{1}{2} p^{\dagger} p + \frac{1}{2(g \chi)^{2}}
\left[ (G_{\rm rad}^{1})^{2} + (G_{\rm rad}^{2})^{2} \right]
+ V(\chi ^{2}) + \frac{1}{2} (\vec{\nabla} \chi)^{2}
\nonumber \\
& & + \frac{1}{2} \left(g \chi \right)^{2} \left[
(\vec{A}^{\,1})^{2} + (\vec{A}^{\,2})^{2} \right]
+ \frac{1}{2} \left( \vec{E}^{a} \vec{E}^{a} + \vec{B}^{a}
\vec{B}^{a} \right)  \ .
\label{3.14}
\end{eqnarray}
The residual Gauss law (\ref{3.11}) takes on the simpler, abelian form
\begin{equation}
G_{\rm rad}^{3} |\ \rangle = - (\vec{\nabla} \vec{E}^{3}
+ g \epsilon^{3bc} \vec{A}^{\,b} \vec{E}^{c} ) |
\ \rangle = 0
\label{3.15}
\end{equation}
and can therefore be further resolved by one of the standard choices
of representations for QED \cite{LNOT12}.
We shall perform this reduction shortly for the particular case of
the Coulomb gauge.

If we approximate $\chi$, the modulus of the matter field, by its vacuum
expectation value, the quadratic part of the
Hamiltonian density for the radiation field derived from (\ref{3.14})
becomes
\begin{equation}
{\cal H}_0  =
\frac{1}{2} \sum_{a=1,2} \left\{ \frac{(\vec{\nabla}
\vec{E}^{a})^{2}}{(g\langle \chi \rangle )^2 }
+ (\vec{E}^{a})^{2} + (\vec{\nabla} \times \vec{A}^{\,a} )^{2}
+ (g \langle \chi \rangle)^2 (\vec{A}^{\,a})^2 \right\}
 +  \frac{1}{2} (\vec{E}^{3})^{2} +
\frac{1}{2} ( \vec{\nabla} \times \vec{A}^{\,3} )^{2}  \ .
\label{3.16}
\end{equation}
As expected, the vector field
$\vec{A}^{\,3}$ stays massless, whereas $\vec{A}^{\,1}$ and $\vec{A}^{\,2}$
acquire a mass $m=g\langle \chi \rangle$ in the same way as in
the abelian Higgs model.
This latter fact can be most easily seen by performing an additional
Bogoliubov transformation, cf. Ref. \cite{LNOT12}. The degrees of
freedom at this stage of gauge fixing are: A neutral
scalar Higgs field $\chi$,
a massless, neutral photon field $\vec{A}^3$, and two massive, electrically
charged vector fields, $\vec{W}^{\pm}=\frac{1}{\sqrt{2}}(\vec{A}^1 \mp
i \vec{A}^2)$, analoguous to the $W$-bosons in the standard model.

The residual Gauss law (\ref{3.15})
can be used for instance to eliminate
the longitudinal degrees of freedom of the photon
field $\vec{A}^{\,3}$.
Since this part of the gauge fixing procedure follows closely the QED case
\cite{LNOT12}, we can be rather sketchy.
One first solves the Gauss law with respect to the longitudinal part
of $\vec{E}^3$,
\begin{equation}
\vec{E}^3|\ \rangle = \left( \vec{E}^{3,tr} + \frac{1}{V}\vec{E}^{3,0}
+ \vec{\eta}^{\,3} \right) | \ \rangle \ .
\label{3.17}
\end{equation}
Here, $\vec{\eta}^{\,3}$ is the
longitudinal, electrostatic field
\begin{equation}
\vec{\eta}^{\,3} (\vec{x}) = g \vec{\nabla} \int d^{3}y
D(\vec{x}- \vec{y} ) \rho^{3}_{\rm rad}(\vec{y})
\label{3.18}
\end{equation}
expressed in terms of the
periodic Green's function of the Laplacian,
\begin{equation}
D(\vec{z}) = - \frac{1}{V} \sum_{n \neq 0} \frac{1}{p_{n}^{2}}
e^{i \vec{p}_{n} \vec{z}} \ ,
\ \ \ \ \ \ \Delta D(\vec{z}) = \delta^{(3)}(\vec{z})- \frac{1}{V} \ .
\label{3.19}
\end{equation}
This leaves a global constraint, the neutrality condition
\begin{equation}
Q^{3}_{\rm rad}|\ \rangle = 0 \ ,
\label{3.20}
\end{equation}
characteristic for the torus. The UGFT which will eliminate the
longitudinal part of the conjugate field $\vec{A}^{\,3}$ can easily
be found,
\begin{equation}
{\cal U}_{\alpha} = \exp \left\{ -ig \int d^{3}x \rho^{3}_{\rm rad}
(\vec{x}) \alpha^{3}(\vec{x}) \right\} \ ,
\label{3.21}
\end{equation}
with
\begin{equation}
\alpha^{3}(\vec{x}) = \int d^{3}y D(\vec{x}- \vec{y})
\vec{\nabla}  \vec{A}^{\,3}(\vec{y}) \ .
\label{3.22}
\end{equation}
As a result of this 2nd unitary transformation, the Hamiltonian density
$\tilde{\cal H}' = {\cal U}_{\alpha} \tilde{\cal H} \, {\cal U}_{\alpha}^
{\dagger} $, projected onto the physical space,
is solely expressed in terms of unconstrained variables as follows,
\begin{eqnarray}
\tilde{\cal H}' & = & \frac{1}{2} p^{\dagger} p +
\frac{1}{2(g \chi )^{2}}
\left[ (G_{\rm rad}^{'1})^{2} + (G_{\rm rad}^{'2})^{2} \right] + V(\chi^{2})
+ \frac{1}{2} (\vec{\nabla} \chi )^{2}
\nonumber \\
& &
+ \frac{1}{2} (g \chi )^{2} \left[ (\vec{A}^{\,1})^{2} +
(\vec{A}^{\,2})^{2} \right]
+ \frac{1}{2} \left[ \vec{E}^{\,'a}\vec{E}^{\,'a} +
\vec{B}^{\,'a} \vec{B}^{\,'a} \right] \ .
\label{3.23}
\end{eqnarray}
Here, the primed quantities differ from the standard
ones by the substitutions
\begin{eqnarray}
\vec{A}^{\,3} & \to & \vec{A}^{\,'3} = \vec{A}^{\,3,tr} + \vec{A}^{\,3,0}
 \ , \nonumber \\
\vec{E}^3 & \to & \vec{E}^{\,'3} = \vec{E}^{3,tr} + \frac{1}{V}
\vec{E}^{3,0} + \vec{\eta}^{\,3} \ ,
\label{3.24}
\end{eqnarray}
confirming that we have succeeded in eliminating the longitudinal
photon degrees of freedom from eq. (\ref{3.14}).

The Hamiltonian (\ref{3.23}) in a particular combination of unitary and
Coulomb gauge is the main result so far. We stress once more that we have not
fixed the gauge beyond the Weyl gauge, but have only projected onto
the physical space and transformed
the Hamiltonian to a new representation which will be referred to as
``unitary/Coulomb gauge
representation''. Needless to say, eq. (\ref{3.23})
could have been derived in a variety of ways. Our quantum
mechanical method has some advantages which will be exploited
in the following sections.

\setcounter{equation}{0}
\section
{The 't Hooft tensor and other gauge invariant operators}

Interest in the Georgi Glashow model stems primarily from the existence
of magnetic monopoles, which were originally discovered by 't Hooft and
Polyakov \cite{Hooft74,Polyakov}. In this context,
't Hooft has proposed a gauge invariant
(Lorentz) tensor closely related to the abelian magnetic field.
In our notation, it is given by
\begin{equation}
{\cal F}_{\mu \nu} = \frac{1}{\chi} \phi^a F_{\mu \nu}^{a}
- \frac{1}{g \chi^3}
\epsilon_{abc} \phi^a (D_{\mu} \phi)^b (D_{\nu} \phi)^c \ .
\label{4.1}
\end{equation}
In the unitary gauge where $\phi^a = \chi \delta_{a3}$, this
tensor reduces to
\begin{equation}
{\cal F}_{\mu \nu} = \partial_{\mu} A_{\nu}^{3} -
\partial_{\nu}A_{\mu}^{3} \ ,
\label{4.2}
\end{equation}
i.e., its spatial components describe the abelian magnetic field.
Eq. (\ref{4.1}) can serve as starting point for
the discussion of 't Hooft-Polyakov monopoles \cite{Hooft74,Polyakov,Rossi}
and
has been found useful also in other instances, notably in the context
of lattice gauge theory where the gauge is in general not fixed.
We follow the common practice to refer to
${\cal F}_{\mu \nu}$ of eq. (\ref{4.1}) as 't Hooft tensor.

Let us first try to locate this well-known object
within our
framework.
In the unitary gauge representation corresponding to the Hamiltonian
(\ref{3.23}), the abelian magnetic field (not to be confused with
eq. (\ref{3.2})) is
given by
\begin{equation}
B_{i}^{3} := \epsilon_{ijk} \partial_{j} A_{k}^{3} = \frac{1}{2}
\epsilon_{ijk} \left( F_{jk} + ig [A_{j},A_{k}] \right)^{3} \ .
\label{4.3}
\end{equation}
We transform this expression backwards to the original Weyl
gauge by reverting the UGFT's.
The second transformation ${\cal U}_{\alpha}$ (cf. eq. (\ref{3.21}))
does not affect $\vec{B}^3$, therefore only ${\cal U}_{\Delta}$
(\ref{3.13}) enters,
\begin{equation}
{\cal U}_{\Delta}^{\dagger} B_{i}^{3}\, {\cal U}_{\Delta} =
\frac{1}{2} \epsilon_{ijk} \left\{ \left( e^{-ig \Delta} F_{jk}
e^{ig\Delta} \right)^{3}
+ ig \, {\cal U}_{\Delta}^{\dagger} \left( [A_{j}, A_{k}]
\right)^{3} {\cal U}_{\Delta} \right\} \ .
\label{4.4}
\end{equation}
With $\hat{\phi}$ as introduced in eq. (\ref{3.12}), the first term is
obviously
\begin{equation}
\epsilon_{ijk}  \mbox{tr} \left(  \hat{\phi} F_{jk} \right) \ .
\label{4.5}
\end{equation}
The 2nd term can be transformed into a more familiar form as follows:
Starting from
\begin{equation}
D_{j} \hat{\phi}  = -ig e^{ig \Delta} \left[ e^{-ig \Delta} \left( A_{j}
+ \frac{i}{g} \partial_{j} \right) e^{ig \Delta} , \frac{\sigma^{3}}{2}
\right] e^{-ig \Delta}
\label{4.6}
\end{equation}
and using the following matrix identity which holds for $\hat{C} \hat{C} = 1$,
\begin{equation}
\mbox{tr} \left( \hat{C} [[A, \hat{C}],[B,\hat{C}] ] \right)
= 4 \,\mbox{tr} \left(
\hat{C} [B,A] \right) \ ,
\label{4.7}
\end{equation}
we can show that
\begin{equation}
\mbox{tr} \left( \hat{\phi} [ D_{j} \hat{\phi}, D_{k} \hat{\phi} ]
\right)  =
\frac{1}{2}g^2 {\cal U}_{\Delta}^{\dagger} \left(
[A_{j}, A_{k}] \right)^{3} {\cal U}_{\Delta} \ .
\label{4.8}
\end{equation}
Hence,
\begin{equation}
{\cal U}_{\Delta}^{\dagger} B_{i}^{3} {\cal U}_{\Delta} =
\frac{1}{2} \epsilon_{ijk} {\cal F}_{jk} \ ,
\label{4.9}
\end{equation}
with
\begin{equation}
{\cal F}_{jk} = 2 \, \mbox{tr} \left( \hat{\phi} F_{jk} +
\frac{i}{g} \hat{\phi}
[D_{j} \hat{\phi}, D_{k} \hat{\phi} ] \right) \ ,
\label{4.10}
\end{equation}
in agreement with the spatial components of the 't Hooft tensor
(\ref{4.1}).
We note in passing
that ${\cal F}_{jk}$ can be rewritten
in a simpler way such that no terms quadratic in $A$
appear \cite{Arafune}.
After some algebra, one finds
\begin{equation}
{\cal F}_{jk} = 2 \, \mbox{tr} \left( \partial_{j} ( \hat{\phi} A_{k}) -
 \partial_{k} ( \hat{\phi} A_{j})+  \frac{i}{g} \hat{\phi}
[\partial_{j} \hat{\phi} , \partial_{k} \hat{\phi} ] \right) \ .
\label{4.11}
\end{equation}
The gauge invariance of the tensor (\ref{4.11}) is no longer manifest,
but it is more convenient for practical applications.
By way of example, it can be used to
derive the well-known
connection between zeros of the scalar field and positions
of magnetic monopoles \cite{Arafune}. This connection
becomes essential
if one is interested in deviations from
the Higgs phase, in particular the transition to the confining
phase. Since the technique used here has some new twist, yet magnetic
monopoles are not the main subject of this work, we have deferred the
corresponding derivation to the appendix.

The 't Hooft tensor was only one example of how to exhibit the
gauge invariant meaning of certain operators appearing in the gauge fixed
formulation. If we want to translate any operator ${\cal O}$ back into the
original Weyl gauge variables, all we have to do is to perform the inverse
UGFT,
\begin{equation}
{\cal O} \to  {\cal U}^{\dagger} {\cal O} {\cal U} \qquad \ \ \ \ \left(
{\cal U} := {\cal U}_{\alpha} {\cal U}_{\Delta} \right) \ .
\label{4.12}
\end{equation}
It is instructive to transform backwards
the charge density $\rho^3_{\rm rad}$. A gauge invariant definition
of the electric charge density is then seen to be
\begin{equation}
\rho_{\rm rad}^3 \to {\cal U}^{\dagger} \rho_{\rm rad}^3 {\cal U}
= \frac{1}{g} ( \vec{D}\hat{\phi} )^a \vec{E}^a =\frac{1}{g}
{\cal G}_{\rm rad}^a \hat\phi^a \ . \label{4.13}
\end{equation}
Using the Gauss law, this can equivalently be expressed as a divergence,
\begin{equation}
{\cal U}^{\dagger} \rho_{\rm rad}^3 {\cal U} = \frac{1}{g}
\vec{\nabla} (\vec{E}^a \hat{\phi}^a ) \ .
\label{4.14}
\end{equation}
Eq. (\ref{4.14})
is just the abelian Gauss law of electromagnetism, emerging
as a relation between gauge invariant electric field and charge
density from a non-abelian Higgs model. Similarly, we could
apply the recipe (\ref{4.12}) to the physical field variables
$\vec{A}^a$, etc.

Gauge invariant, ``composite'' fields and operators of similar type
as the 't Hooft tensor have appeared
repeatedly in the literature and have been found quite
useful. Thus for instance,
Witten \cite{Witten} has invoked the gauge invariant
electric charge operator (\ref{4.13})
together with the corresponding magnetic charge
in a study of dyons, particles that carry both electric
and magnetic charge.
A lot of effort has been spent on a
``complementarity
principle'', trying to explain intuitively the finding that there
is no sharp phase boundary between Higgs and confined phase in the fundamental
Higgs model \cite{Strocchi}--\cite{Dimopoulos}. In these references,
gauge invariant, ``composite'' fields play an important role
for clarifying conceptual issues.
They may also serve to relate
gauge fixed formulations to lattice gauge theories.
Below, we will apply similar techniques to understand the origin of
global, residual symmetries in non-abelian Higgs models.

\setcounter{equation}{0}
\section{Displacement symmetry of the Georgi Glashow model}

In QED, the photon
can be identified with the Goldstone boson of the spontaneously broken
displacement symmetry. If a massless vector particle belongs to
the spectrum of the Georgi Glashow model, one would expect
a similar mechanism to be at work here as well. Inspection of the
Hamiltonian (\ref{3.23}) in the unitary/Coulomb gauge representation
indeed reveals residual symmetries.
These symmetries are of two types: Global rotations in internal space
generated by charge operators, and abelian, displacement type
symmetries generated by a displacement vector.

Let us first consider the residual global gauge transformations.
The Hamiltonian (\ref{3.23}) is invariant under arbitrary rotations
around the 3-axis, as well as rotations by $\pi$ about
any axis perpendicular to the 3-direction (or products hereof) --
the normalizer of the SO(2) subgroup, $N$(SO(2))
\cite{Ovrut}. This particular symmetry is easy to verify, since the
subtraction
of the longitudinal, neutral fields does not interfere with the
structure in internal space. All one has to do is define the charge
$Q_{\rm rad}^a$ with $\vec{A}^3, \vec{E}^3$ replaced by
$\vec{A}^3-\vec{A}^{3,\ell}$, $\vec{E}^3-\vec{E}^{3,\ell}$, respectively.
The global $N$(SO(2)) residual gauge group is somewhat misleading however,
since we still have an unresolved constraint, the neutrality
condition (\ref{3.20}).
It implies that global rotations about the 3-axis
are reduced to ${\bf 1}$
in the physical space. Rotations about an axis in the (1,2)-plane
by $\pi$,
\begin{equation}
\Omega_{\vec{n}_{\bot}} = \exp \left\{ -i\pi \left( n_1 Q_{\rm rad}^1
+ n_2 Q_{\rm rad}^2 \right) \right\} \ ,  \qquad
\left( n_1^2+n_2^2=1 \right) \ ,
\label{5.1}
\end{equation}
do not lead out of the physical space, since
\begin{equation}
Q_{\rm rad}^3 \Omega_{\vec{n}_{\bot}} |\ \rangle =
- \Omega_{\vec{n}_{\bot}} Q_{\rm rad}^3 |\ \rangle = 0 \ .
\label{5.2}
\end{equation}
Thus, in the physical space, the group $N$(SO(2)) will be reduced to
$N$(SO(2))/SO(2) $\simeq Z_2$. Physically, this discrete,
global symmetry is
just the ordinary charge conjugation symmetry, but now for a model where
electromagnetism is embedded in a SU(2) gauge theory in a non-trivial
way.

The second type of symmetry involves
linearly $\vec{x}$ dependent gauge functions (``displacement symmetry'').
Since it is not generated by $Q_{\rm rad}^3$ (but commutes with it), it
survives in the
physical sector as a genuine symmetry. It is represented by the operator
\begin{equation}
\Omega_{\vec{n}} = \exp \left\{-i \frac{2\pi}{gL} \vec{n} \vec{\cal D}\right\}
\label{5.3}
\end{equation}
with the displacement vector
\begin{equation}
\vec{\cal D}  =
 \int d^3 x \left( \vec{E}^3 + g \vec{x}
\rho^3_{\rm rad} \right)  \ .
\label{5.4}
\end{equation}
Just as in electrodynamics,
$\Omega_{\vec{n}}$ shifts the photon field by a constant
vector and rotates the phase of the electrically charged fields by
an angle with
linear $\vec{x}$-dependence (or, in more physical
terms, shifts the momenta of all charged particles by a constant),
\begin{eqnarray}
\Omega_{\vec{n}} \vec{A}^3(\vec{x}) \Omega_{\vec{n}}^{\dagger} =
\vec{A}^3
(\vec{x}) + \frac{2\pi}{gL} \vec{n}          \ ,
\nonumber \\
\Omega_{\vec{n}} \vec{W}^{\pm}(\vec{x}) \Omega_{\vec{n}}^{\dagger}
= e^{\pm i \frac{2\pi}{L}\vec{n}\vec{x}} \vec{W}^{\pm}(\vec{x}) \ .
\label{5.5}
\end{eqnarray}
This displacement symmetry which emerges in the unitary/Coulomb gauge
representation of the Georgi Glashow model is strongly reminiscent of QED.
The analogy to QED is not perfect, though: In QED,
the displacement vector
is not affected by the UGFT leading to the
Coulomb gauge
representation. Hence one can identify the
displacements with a certain
kind of gauge transformations at the level of the Weyl gauge Hamiltonian,
namely the
topologically non-trivial, ``large" gauge transformations for the U(1) theory
on the torus \cite{LNOT12}. In the Georgi Glashow model,
the displacement vector is affected in the process of gauge fixing.
In order to exhibit its gauge invariant meaning,
we follow the procedure
outlined in the preceding section and simply apply the inverse UGFT to
expression (\ref{5.4}).  It is sufficient to
consider ${\cal U}_{\Delta}$, since ${\cal U}_{\alpha}$
commutes with $\vec{\cal D}$.
We then obtain the manifestly gauge invariant result
\begin{equation}
\vec{\cal D} \to {\cal U}_{\Delta}^{\dagger} \vec{\cal D} {\cal U}_{\Delta} =
\int d^3x  {\cal G}_{\rm rad}^a \vec{x} \hat{\phi}^a \ . \label{5.6}
\end{equation}
Correspondingly,
the symmetry operator $\Omega_{\vec{n}}$, eq. (\ref{5.3}), can be associated
with the following unitary operator at the level of the Weyl gauge,
\begin{equation}
\Omega_{\vec{n}} \to {\cal U}_{\Delta}^{\dagger} \Omega_{\vec{n}}
{\cal U}_{\Delta} =
\exp \left\{ -i \frac{2\pi}{gL}
\int d^3x  {\cal G}_{\rm rad}^a (\vec{n}\vec{x}) \hat{\phi}^a
\right\}\ .
\label{5.7}
\end{equation}
Clearly, this is not an ordinary gauge transformation.
It represents
a gauge transformation of the radiation variables with a gauge function
depending on the direction of the matter field in the internal space:
The matter field dictates the local ``rotation axis''. A second difference
to the familiar QED case shows up if we try to evaluate $\Omega_{\vec{n}}
H \Omega_{\vec{n}}^{\dagger}$, with $H$ the Weyl gauge Hamiltonian
(obtained by integrating (\ref{3.1}) over $d^3x$) and
$\Omega_{\vec{n}}$ as given in eq. (\ref{5.7}).
In QED, the correponding operators $\Omega_{\vec{n}}$ and $H$ commute even
in the extended Hilbert space where the Gauss law is not enforced,
since $\Omega_{\vec{n}}$ is a special
kind of gauge transformation. In the Georgi Glashow model, the matter
field kinetic energy
$\pi^a \pi^a /2$ fails to commute with $\Omega_{\vec{n}}$ due to
the $\hat{\phi}^a$ dependence of the latter. There is
no contradiction with the fact that the displacement symmetry is an
exact symmetry of the Hamiltonian (\ref{3.23}),
because there we restricted
ourselves to the physical space. In the Weyl gauge,
what happens can be understood as follows:
Separating the kinetic energy of the scalar field
into radial and angular parts ($p$ is the same operator which has been
used in Sect. 4)
\begin{equation}
\pi^a \pi^a = p^{\dagger} p + \frac{1}{\chi^2} \rho_{\rm matt}^a
\rho_{\rm matt}^a
\label{5.8}
\end{equation}
and introducing the Gauss law operator (\ref{3.3}),
we get the identity
\begin{equation}
\pi^a \pi^a = p^{\dagger} p + \frac{1}{(g\chi)^2} \left(G^a-
G_{\rm rad}^a\right)
\left(G^a-G_{\rm rad}^a\right) \ .
\label{5.9}
\end{equation}
It entails a
corresponding decomposition of the Weyl gauge Hamiltonian density,
\begin{equation}
{\cal H} = {\cal H}' + \frac{1}{2(g\chi)^2} \left(G^aG^a
- G^a_{\rm rad} G^a -
G^a G_{\rm rad}^a \right) \ ,
\label{5.10}
\end{equation}
where ${\cal H}'$ differs from eq. (\ref{3.1}) by the replacement
\begin{equation}
\pi^a \pi^a \to p^{\dagger} p + \frac{1}{(g\chi)^2} G_{\rm rad}^a
G_{\rm rad}^a \ ,
\label{5.11}
\end{equation}
and the second term in (\ref{5.10}) vanishes in the physical sector.
Now, it is easy to check that $H'= \int d^3x {\cal H}'$ is invariant
under displacements (\ref{5.7}),
\begin{equation}
\Omega_{\vec{n}} H' \Omega_{\vec{n}}^{\dagger} = H' \ ,
\label{5.12}
\end{equation}
in spite of the fact that
$\Omega_{\vec{n}}$ is not a gauge transformation. The reason is
of course that the scalar field is invariant
under rotations about its own direction.
The difference $H-H'$ does not commute with $\Omega_{\vec{n}}$, but since
it vanishes in the physical sector,
this is of no concern to us. (Note that the symmetry could be trivially
extended to the large Hilbert space by letting $\Omega_{\vec{n}}$
act only in the physical space and using a {\bf 1} in the unphysical
space; however, then it would loose its simple closed form,
eq. (\ref{5.7}).)
Here, we see a significant difference between the
abelian and non-abelian cases. In particular, it seems that
the displacement symmetry of the Georgi Glashow model
is not related in any simple way to ``large'' gauge transformations;
as a matter of fact,
it does not correspond to a gauge transformation at all.

Summarizing, in the unitary/Coulomb
gauge representation of the Georgi Glashow model, we observe superficially
the same kind
of displacement symmetry as in QED. There is no doubt that this
symmetry is spontaneously broken in the phase which supports
massless photons. In contrast to QED,
the displacement operator is modified in the process of fixing
the gauge. Tracing it backwards, we find its gauge invariant meaning
in the same manner as one can derive the 't Hooft tensor by asking for a
gauge invariant definition of the abelian magnetic field. The gauge
invariant displacement operator (\ref{5.6}) describes certain rotations
of the gauge field in internal space about an
axis defined locally by the matter field. Since this is only
well defined if the modulus of the Higgs field does not vanish, one
can understand at once why the symmetry breaking
in SU(2) Yang
Mills theory requires the presence of a scalar matter field
with a non-vanishing condensate.
By contrast, in QED, symmetry breaking already occurs in the
free theory; here, a scalar field with non-vanishing expectation
value has the opposite effect of restoring the displacement symmetry
\cite{LNOT12}.

We conclude this section with a few remarks on topological issues.
The displacement symmetry of QED on a torus coincides with
the topologically non-trivial gauge transformations. These in turn owe their
existence to the fact that the homotopy group for mappings from $T^3$
to $S^1$ is non-trivial and isomorphic to $Z^3$. In the Georgi Glashow
model, we have also found a displacement symmetry when
working in the unitary/Coulomb representation. However, there seems
to be no direct
relationship with topologically non-trivial gauge transformations.
The corresponding
groups do not match -- the homotopy group for $T^3 \to \mbox{SO(3)}$ is
$Z_2^3 \times Z$, cf. \cite{Baal}, whereas we observe a combination
of $Z_2$ (charge conjugation) with $Z^3$ (displacements).
Spontaneous breakdown of the group of ``large'' gauge transformations
in the non-abelian case
cannot be held reponsible for the existence
of massless vector particles; at best, if the $Z$ symmetry would become
continuous in the limit $L \to \infty$,
it could explain a massless scalar.
We see no trace of the homotopically non-trivial gauge transformations.
Presumably, they are reduced to {\bf 1} in the Higgs phase,
as in the abelian case.

\setcounter{equation}{0}
\section{Fundamental Higgs fields and standard model}

Continuing our inventory of the most important Higgs models, we now turn
to fundamental Higgs fields.
Let us first consider a pure SU(2) Higgs model with scalar fields in the
fundamental representation, with
the Hamiltonian density
\begin{equation}
{\cal H}  =  \frac{1}{2} ( \vec{E}^a \vec{E}^a + \vec{B}^a \vec{B}^a)
+ \pi^{\dagger} \pi + ( \vec{D} \varphi )^{\dagger}
( \vec{D} \varphi ) + V ( \varphi^{\dagger} \varphi )  \ .
\label{6.1}
\end{equation}
$\vec{D}$ now stands for the
covariant derivative in the fundamental representation,
\begin{equation}
\vec{D} = \vec{\nabla} -  \frac{i}{2} g
\vec{A}^a \sigma^a   \ .
\label{6.2}
\end{equation}
The fundamental Higgs field consists of a complex doublet,
\begin{equation}
\varphi = \frac{1}{\sqrt{2}} \left(
\begin{array}{c} \varphi_1 + i \varphi_2 \\ \varphi_3 + i \varphi_4
\end{array} \right) \ ,
\label{6.3}
\end{equation}
with a corresponding expression for the canonical momenta $\pi$.
For later use, we decompose $\varphi$ into a hermitian field $\chi$
and a SU(2) matrix,
\begin{equation}
\varphi = \frac{1}{\sqrt{2}} e^{ig\Delta} \left( \begin{array}{c}
0 \\ \chi \end{array} \right)  = \frac{1}{\sqrt{2}} \chi \hat{\varphi} \ .
\label{6.4}
\end{equation}
The SU(2) matrix can be expressed in terms of the components of
$\hat{\varphi}$ as
\begin{equation}
e^{ig\Delta} = \left( \begin{array}{rr} \hat{\varphi}_3-i\hat{\varphi}_4 &
\hat{\varphi}_1+i\hat{\varphi}_2 \\ -\hat{\varphi}_1+i\hat{\varphi}_2 &
\hat{\varphi}_3+i\hat{\varphi}_4 \end{array} \right)
\label{6.5}
\end{equation}
The Gauss law has the same form as in eq. (\ref{3.3}), with
the matter SU(2) charge density now given by
\begin{eqnarray}
\rho_{\rm matt}^a &=& \frac{i}{2} ( \varphi^{\dagger} \sigma^a \pi
- \pi^{\dagger} \sigma^a \varphi )   \nonumber  \\
 &=& \frac{1}{2} \chi \sum_{i=1}^4 v^a_i  \pi_i :=
 \frac{1}{2}\chi (v^a,\pi)\  \qquad (a=1,2,3) \ .
\label{6.6}
\end{eqnarray}
Here, we have introduced
real, 4-component vectors $\{v^a,v^4\}$ which form the following complete,
orthonormal set
\begin{equation}
(v^1,v^2,v^3,v^4)= \left( \begin{array}{rrrr}
\hat{\varphi}_4 & -\hat{\varphi}_3 & \hat{\varphi}_2 & \hat\varphi_1 \\
-\hat{\varphi}_3 & -\hat{\varphi}_4 & -\hat{\varphi}_1 & \hat\varphi_2 \\
\hat{\varphi}_2 & \hat{\varphi}_1 & -\hat{\varphi}_4 & \hat\varphi_3 \\
-\hat{\varphi}_1 & \hat{\varphi}_2 & \hat{\varphi}_3& \hat\varphi_4
\end{array}    \right) \ .
\label{6.9}
\end{equation}
After these preparations, the
resolution of Gauss's law
\begin{equation}
\left(G_{\rm rad}^a + \frac{1}{2}g \chi (v^a, \pi )\right)|\ \rangle = 0
\label{6.10}
\end{equation}
is no more difficult than in the case of the abelian Higgs model.
Denoting the unconstrained, radial momentum $(v^4,\pi)$
once more by $p$, we have
\begin{equation}
\pi_i |\  \rangle = \left( -\frac{2}{g\chi} \sum_{a=1}^3 (v_i^a
G_{\rm rad}^a )  + v_i^4 p  \right) |\ \rangle   \ ,
\label{6.11}
\end{equation}
and consequently the
matter kinetic energy in the physical space becomes
\begin{equation}
\langle \ |  \pi^{\dagger}\pi | \ \rangle =
\langle \ | \left( \frac{1}{2} p^{\dagger}p + \frac{2}{(g\chi)^2}
G^a_{\rm rad} G^a_{\rm rad} \right) |\ \rangle      \ .
\label{6.12}
\end{equation}
The expression for the unitary operator
${\cal U}_{\Delta}$ is unchanged as compared to eq. (\ref{3.13}),
provided we take $\Delta$ from eq. (\ref{6.4}).
The transformed Hamiltonian in the physical space is then
found to be
\begin{equation}
\tilde{\cal H}  =
 \frac{1}{2} ( \vec{E}^a \vec{E}^a + \vec{B}^a \vec{B}^a)
+ \frac{1}{2} p^{\dagger}p + \frac{2}{(g\chi)^2} G^a_{\rm rad}
G^a_{\rm rad}
 + \frac{1}{2} (\vec{\nabla}\chi)^2
+ \frac{g^2}{8} \chi^2 \vec{A}^a \vec{A}^a + V(\chi^2/2) \ .
\label{6.13}
\end{equation}
Just as in the unitary gauge representation of the abelian Higgs model,
the constraints have been completely resolved. If one now replaces
$\chi$ by its $c$-number part $\langle \chi \rangle$ and inspects the
quadratic part of
the Hamiltonian (\ref{6.13}), one finds that all three
vector bosons acquire the same mass $g\langle \chi \rangle /2$. This
degeneracy reflects
a global SU(2) residual symmetry which
$\tilde{\cal H}$ still possesses,
generated by the charges $gQ_{\rm rad}^a$.

In order to understand the origin of this symmetry, we use
the same strategy as above and transform the generators
backwards via the inverse UGFT to
find their gauge invariant form,
\begin{equation}
gQ^a_{\rm rad} \to {\cal U}_{\Delta}^{\dagger}gQ_{\rm rad}^a
{\cal U}_{\Delta} = \int d^3x \left(e^{-ig\Delta} G_{\rm rad}
e^{ig\Delta} \right)^a \ .
\label{6.14}
\end{equation}
At the level of the
Weyl gauge, these charges generate gauge transformations of the
radiation field with global rotation angle, but rotation axis depending
locally on the direction of the Higgs field. Hence, one can view
this symmetry as another example of the
phenomenon discussed above in the Georgi Glashow model, where the
presence of the Higgs field gives rise to new, global symmetries
in the physical space, remnants of field-dependent gauge transformations
in the classical theory.
In our formulation,
there are many equivalent ways of writing down such gauge invariant,
``composite''
operators, provided one makes use of the Gauss law. In the case at hand,
a more illuminating
representation results if we trade $G^a_{\rm rad}$
for the matter charge density and express the generator
entirely in terms of Higgs field variables. This requires
some further formal tools. With the help of the orthogonal matrix
\begin{equation}
R^{ab}= \frac{1}{2} \mbox{Tr} \left( e^{ig\Delta} \sigma^a
e^{-ig\Delta} \sigma^b \right) \ ,
\label{6.15}
\end{equation}
we define the SU(2) charge density in the ``intrinsic frame'' via
\begin{equation}
\tilde{\rho}^a_{\rm matt} = R^{ab} \rho^b_{\rm matt} \ .
\label{6.16}
\end{equation}
Correspondingly, the space integral of $\tilde{\rho}_{\rm matt}^a$ will
be denoted by $\tilde{Q}_{\rm matt}^a$.
As is well known, the $\rho_{\rm matt}^a$ and $\tilde{\rho}_{\rm matt}^a$
can be interpreted
either as generators of local left- or right-rotations of $e^{ig\Delta}$,
or as components of local angular momentum operators in the laboratory
or body-fixed frames, with the appropriate commutation relations
\cite{Kogut}.
In the physical space, using Gauss's law, we can
then replace the generator of
the residual symmetry (\ref{6.14}) by the simpler expression
\begin{equation}
{\cal U}_{\Delta}^{\dagger}gQ_{\rm rad}^a {\cal U}_{\Delta} | \ \rangle
= \int d^3x R^{ab}G_{\rm rad}^b | \ \rangle
= -g \tilde{Q}^a_{\rm matt}| \ \rangle \ .
\label{6.17}
\end{equation}
Thus the global symmetry which survives in the physical space
and acts on radiation fields can be interpreted as (inverse) global SU(2)
transformation on the matter field, but in the intrinsic frame.
It is instructive to compare these findings with the symmetries of the
ungauged fundamental
Higgs model: There, one starts out with a larger O(4)
symmetry, which gets broken spontaneously down to O(3), due to the assumed
form of the potential. Equivalently, one might describe this
situation by saying that a SU(2)$\times$SU(2) symmetry (left and
right rotations of $e^{ig\Delta}$) is broken down to SU(2) (the
analogy between this pattern and breaking of chiral symmetry in
QCD has recently been noted \cite{Leutwyler}). When gauging this
model, the left rotations are gauged and disappear, whereas the
ungauged right rotations survive and are inherited by the gauge
field, in the unitary gauge. This is exactly what
eq. (\ref{6.17}) is telling us.

Let us now extend this model to the bosonic sector of the GWS theory,
by considering a
U(1)$\times$SU(2) Higgs model with fundamental scalar field.
It is easy to augment the preceding model by a local U(1) gauge
group. The
Weyl gauge Hamiltonian density becomes
\begin{equation}
{\cal H}   =   \frac{1}{2} ( \vec{E}^a \vec{E}^a + \vec{B}^a \vec{B}^a)
+ \frac{1}{2} ( \vec{\cal E}^{\,2} + \vec{\cal B}^{\,2} )
+ \pi^{\dagger} \pi + ( \vec{D} \varphi )^{\dagger}
( \vec{D} \varphi ) + V ( \varphi^{\dagger} \varphi )
\label{6.18}
\end{equation}
where $\vec{\cal B}= \vec{\nabla} \times
\vec{\cal A}$, and
the covariant derivative should be interpreted this time as
\begin{equation}
\vec{D} = \vec{\nabla} -  \frac{i}{2} ( g'\vec{\cal A} + g
\vec{A}^a \sigma^a )  \ .
\label{6.19}
\end{equation}
In addition to the SU(2) Gauss law (\ref{3.3}), we must impose the
U(1) Gauss law onto the physical states,
\begin{equation}
(- \vec{\nabla} \vec{\cal E}  + g' \rho^0_{\rm matt} )|\ \rangle = 0 \ .
\label{6.20}
\end{equation}
$\rho^0_{\rm matt}$ is the U(1) charge density,
\begin{equation}
\rho^0_{\rm matt} = \frac{i}{2} (\varphi^{\dagger} \pi - \pi^{\dagger}
\varphi ) = -\tilde\rho_{\rm matt}^3\ .
\label{6.21}
\end{equation}
Since the U(1) gauge transformations are generated by acting on
the ``angular'' part of the Higgs field $e^{ig\Delta}$
(cf. (\ref{6.4})), it is clear that the corresponding
generator
will be modified during the gauge fixing procedure constructed to
eliminate these angular degrees of freedom. Like the global SU(2)
symmetry discussed previously in the context of the pure SU(2) gauge
group with fundamental Higgs field, the U(1) symmetry generator will
act only on gauge field degrees of freedom, after implementing the
unitary gauge.
We can resolve the SU(2) Gauss law and perform the 1st UGFT exactly as
above. This yields the following
Hamiltonian density in the physical sector,
\begin{eqnarray}
\tilde{\cal H}  & = &
 \frac{1}{2} ( \vec{E}^a \vec{E}^a + \vec{B}^a \vec{B}^a)
 + \frac{1}{2} (\vec{\cal E}^{\,2} + \vec{\cal B}^{\,2})
+ \frac{1}{2} p^{\dagger}p + \frac{2}{(g\chi)^2} G^a_{\rm rad}
G^a_{\rm rad}  \label{6.22} \\
& & + \frac{1}{2} (\vec{\nabla}\chi)^2
+ \frac{g^2}{8} \chi^2 \left[(\vec{A}^{\,1})^2+ (\vec{A}^{\,2})^2\right]
+ \frac{1}{8} \chi^2 (g \vec{A}^{\,3}-g'\vec{\cal A}\,)^2 + V(\chi^2/2) \ .
\nonumber
\end{eqnarray}
If we identify the physical $W^{\pm}$ and $Z$ boson fields as
\begin{equation}
\vec{W}^{\pm} = \frac{1}{\sqrt{2}}(\vec{A}^{\,1} \mp i \vec{A}^{\,2}) \ ,
\qquad \vec{Z} = \cos \theta_W \vec{A}^{\,3} - \sin \theta_W \vec{\cal A} \ ,
\label{6.22a}
\end{equation}
where the Weinberg angle $\theta_W$ is defined as usual,
\begin{equation}
\tan \theta_W = \frac{g'}{g} \ ,
\label{6.22b}
\end{equation}
we can read off eq. (\ref{6.22}) the standard masses of the heavy vector
bosons,
\begin{equation}
M_W = M_Z \cos \theta_W = \frac{1}{2} g \langle \chi \rangle \ .
\label{6.22c}
\end{equation}
The physical photon
field, $\cos \theta_W \vec{\cal A} + \sin \theta_W \vec{A}^{\,3}$, does
not acquire a mass term in the Hamiltonian. Related to this, we
still have the U(1) Gauss
law (\ref{6.20}) which needs to be transformed unitarily as well.
First, we eliminate the matter charge density in (\ref{6.20}) in
favour of
the corresponding radiation field Gauss law operator, using eqs.
(\ref{6.10}), (\ref{6.16}) and (\ref{6.21}).
Next, we perform the unitary transformation
which only modifies $G_{\rm rad}$ via a gauge transformation. Together
with the orthogonality of the matrices $R$, eq. (\ref{6.15}), we find
\begin{equation}
\left(- \vec{\nabla}\vec{\cal E} + \frac{g'}{g} G^3_{\rm rad} \right)|
\ \rangle = 0    \ .
\label{6.28}
\end{equation}
If we identify the electric charge in the standard way,
\begin{equation}
e = \frac{gg'}{\sqrt{g^2+ g^{'2}}} \ ,
\label{6.29}
\end{equation}
and introduce the Weinberg angle (\ref{6.22b}),
we obtain the equivalent form
\begin{equation}
\left( - \vec{\nabla} ( \cos \theta_W \vec{\cal E} + \sin \theta_W
\vec{E}^3) + e \rho^3_{\rm rad} \right) | \ \rangle = 0 \ .
\label{6.30}
\end{equation}
This is the Gauss law of electrodynamics as it appears in (the bosonic
sector of) the GWS model.

We could proceed now and resolve the residual Gauss law (\ref{6.30})
aiming at
the Coulomb gauge representation, as we did for the Georgi Glashow
model. However, this is not necessary in order to discuss the
symmetry aspects which interest us. The displacement symmetry
responsible for the appearance of the massless photon can already
be identified at this level of gauge fixing without difficulty.
The Hamiltonian (\ref{6.22}) is invariant under local U(1)
transformations of the following type,
\begin{equation}
\Omega_{\beta} = \exp \left\{ -i \int d^3 x \left( (\cos \theta_W
\vec{\cal E} + \sin \theta_W \vec{E}^3 )\vec{\nabla} + e \rho^3_
{\rm rad} \right) \beta \right\} \ .
\label{6.31}
\end{equation}
This is of course how electromagnetism manifests itself in the unified
theory. Now, we can argue exactly like in the case of QED \cite{LNOT12}:
The function $e^{ie\beta}$ has to be periodic on the torus, i.e.,
$\beta$ can consist of a periodic part and a linear one of the
form $\frac{2\pi}{eL} \vec{n} \vec{x}$. In the physical sector,
the periodic part is
obliterated by the residual Gauss law. The
linear part gives rise to the displacement symmetry (\ref{5.3})
where now the displacement vector assumes the form
\begin{equation}
\vec{\cal D} = \int d^3x \left( \cos \theta_W \vec{\cal E} + \sin \theta_W
\vec{E}^3 + e \vec{x} \rho^3_{\rm rad} \right) \ .
\label{6.32}
\end{equation}
This expression would not be affected by a further UGFT to the
Coulomb gauge.

We should like to draw the attention to the following difference between
Georgi Glashow and GWS models. In the first case, we have found no
possibility to attribute the displacement symmetry to topologically
non-trivial gauge transformations. In the GWS model,
the situation is again closer to QED in this respect. The
homotopy group for mappings $T^3 \to \mbox{U(1)}\times\mbox{SU(2)}$
trivially possesses a $Z^3$ subgroup due to the U(1) factor of the gauge
group. We proceed to show that the displacement symmetry of the GWS model
is directly related to the ``large'' U(1) gauge transformations. Let
us first carry out explicitly the UGFT for the corresponding symmetry
operator,
\begin{displaymath}
{\cal U}_{\Delta} \exp \left\{ -i\frac{2\pi}{g'L} \int d^3x
\left(\vec{\cal E}\vec{\nabla} + g' \rho_{\rm matt}^0\right)
\vec{n}\vec{x} \right\} {\cal U}_{\Delta}^{\dagger} \  =  \hspace{5.0cm}
\end{displaymath}
\begin{equation}
\hspace{5.0cm}
\exp \left\{ -i \frac{2\pi}{eL} \vec{\cal D} \vec{n} \right\}
\exp \left\{ i \frac{2\pi}{L} \int d^3 x \tilde{\rho}^3_{\rm matt}
\vec{n}\vec{x} \right\}  \ ,
\label{6.34a}
\end{equation}
with the displacement vector $\vec{\cal D}$ as defined in eq. (\ref{6.32}).
Eq. (\ref{6.34a}) can
be verified as follows:
The term involving $\vec{\cal E}$
is trivially unchanged (remember that $e=g' \cos \theta_W$). The U(1)
charge density is most easily transformed if one expresses it first
by the generator of the right-rotations, $\tilde{\rho}_{\rm matt}^3$
(see eq. (\ref{6.21})). Using
\begin{eqnarray}
{\cal U}_{\Delta} \left[ \tilde{\rho}_{\rm matt}^3(\vec{x}),
{\cal U}_{\Delta}^{\dagger}
\right] & = & \frac{1}{g} \int d^3 y \, {\cal G}_{\rm rad}^a(\vec{y})
\left( e^{-ig\Delta(\vec{y})} \left[ \tilde{\rho}_{\rm matt}^3(\vec{x}),
e^{ig\Delta(\vec{y})} \right] \right)^a \nonumber \\
& = & -\frac{1}{g} {\cal G}_{\rm rad}^3(\vec{x})
\label{6.34b}
\end{eqnarray}
and the fact that $e=g\sin \theta_W$, eq. (\ref{6.34a}) then
follows. The r.h.s. of (\ref{6.34a}) can be further simplified
if one restricts oneself to the physical space.
In order to see this,
it is necessary to go back to the SU(2) Gauss law and transform it
unitarily, using similar techniques as in eq. (\ref{6.34b}).
The result,
\begin{equation}
{\cal U}_{\Delta} \left(G^a_{\rm rad} + g \rho^a_{\rm matt} \right)
{\cal U}_{\Delta}^{\dagger} = g \rho_{\rm matt}^a \ ,
\label{6.35}
\end{equation}
shows that the charge densities $\rho^a_{\rm matt}$, and therefore
also the ``intrinsic'' densities $\tilde{\rho}^a_{\rm matt}$
(cf. eq. (\ref{6.16})), annihilate
the physical states in the unitary gauge representation.
The 2nd exponential factor on the r.h.s. of eq. (\ref{6.34a}) thus
reduces to unity in the physical sector.
This completes our proof that
the displacement symmetry of the GWS
model can be attributed to ``large'' gauge transformations, exactly
as in QED.

Obviously, the inclusion of fermions in the standard model will
not modify the general structure of the displacement operator
(\ref{6.32}).
All one has to do is to include the electric charge density of the
fermions into the radiation  charge densities.
One then obtains the relevant symmetry
which is spontaneously broken in the electroweak sector of the standard
model, as testified to an incredible accuracy by the physical photon
($m_{\gamma} < 3 \times 10^{-27} \mbox{eV}$ \cite{Montanet}).
As in the abelian Higgs model, we
see no compelling reason
to assume that any other symmetry, for instance invariance under
gauge transformations of the first kind,
should be spontaneously broken.

\setcounter{equation}{0}
\section{Summary and conclusions}

In the traditional view of the Higgs mechanism, the appearance or
non-appearance of a mass term for the gauge fields is taken as
indicator of gauge symmetry breakdown. Elitzur's theorem states
that only global symmetries can be spontaneously broken. Scattered
through the literature, one finds claims that the photon can be
interpreted as Goldstone boson, related to the breakdown of
invariance under linearly $x$-dependent gauge transformations.

This puzzling situation has incited us to reconsider the symmetry
aspects of the Higgs mechanism in a systematic way. Throughout
this work, our guiding principle was
to eliminate all the redundant degrees of freedom
characteristic for gauge theories. This leaves us with a
formulation which is on the same footing as the formulation
of non-gauge theories. By construction, only global symmetries
can survive such a procedure.
The residual symmetries are genuine symmetries which have
direct impact on the spectrum and other physical properties
of the theory. In particular, the only known mechanism
for the occurence of massless bosons in interacting theories
without constraints
is the Goldstone mechanism. Hence, once we have freed gauge theory
from all the superfluous variables, massless bosons can again be taken
as signals
of spontaneous symmetry breakdown. It is then legitimate
to ask which symmetry is responsible for the
appearance of massless vector bosons (``photons'') in
gauge theories.
If one wishes, one can at any
stage re-introduce redundant variables for heuristic purposes,
and this has helped us to
relate the residual global symmetries to
the underlying gauge symmetry in those cases where this relation was
not obvious.

We do not claim to have found in this manner new results which
do not appear in one form or other somewhere in the literature
devoted to the Higgs mechanism. Nevertheless, we
believe that the strength of our approach is its systematic
character and the fact that all the popular Higgs models are
analyzed in a coherent fashion. We regard our
description as an economical way of projecting out those
symmetries which are physically relevant.

Let us summarize the symmetry
properties of gauged Higgs models as viewed from the unitary gauge.
By this we mean a formulation of gauge theories where the
maximum number of matter degrees of
freedom is eliminated with the help of Gauss's law.
In all cases considered here, the only residual Higgs field is a
scalar gauge singlet, corresponding to the radial
variable ($\chi^2 = \varphi^{\dagger}\varphi$). The Higgs
mechanism is characterized by $\langle \chi \rangle \neq 0$,
a gauge invariant statement and at the same time
a precondition for rendering the unitary gauge non-singular.

\begin{enumerate}

\item {\em U(1) abelian Higgs model}

In the Higgs phase,
no (gauge or other) symmetry is broken, and therefore no
massless particle can appear non-perturbatively -- an unfamiliar
interpretation of the Higgs mechanism, but the natural one
if one uses an ``intrinsic'' approach in terms of physical
variables only.
In the absence of a condensate (i.e., in the Coulomb phase), the displacement
symmetry, a residual gauge symmetry, is spontaneously broken.
Massless photons
appear as a result of the Goldstone theorem. There is one subtlety
which shows that the Goldstone mechanism
can generate vector bosons in gauge theories only \cite{LNOT12}:
A vector symmetry would give rise to three
massless particles, whereas a relativistic massless vector particle
has only two polarization states. The Goldstone
theorem is partly evaded by sending one boson
(the longitudinal photon)
into the unphysical sector of Hilbert space, an
option not available in standard non-gauge theories.

\item {\em SU(2) adjoint Higgs model}

If one eliminates all redundant variables from the Georgi Glashow
model, a displacement type symmetry is again observed which \
-- via spontaneous breakdown -- \ accounts for one
massless vector boson. A discrete residual $Z_2$ symmetry
guarantees in addition equal masses for the two charged vector bosons.
These symmetries are only indirectly related to the original
gauge symmetry and reflect specific features of the adjoint Higgs field:
The displacements are topologically non-trivial U(1) transformations
with the rotation axis defined locally by the direction of the Higgs
field vector. This
remnant of (classical) ``field dependent gauge
transformations'' was uncovered by partly ``unfixing'' the gauge, i.e.
reverting the unitary gauge fixing transformations.

\item {\em SU(2) fundamental Higgs model}

Like in the abelian Higgs model,
no symmetry is broken and consequently no massless boson
appears. A residual global SU(2) symmetry has been identified as
arising from the ungauged part of the SU(2)$\times$SU(2) symmetry
group of the original Higgs field.
In gauge theories, it can be passed on to the radiation field
in the process of resolving Gauss's law, and this is exactly what happens
in the unitary gauge.

\item {\em U(1)$\times$SU(2) fundamental Higgs model}

There is again a spontaneously broken displacement symmetry with a
massless Goldstone photon. As in QED, it can be related to ``large'' U(1)
gauge transformations. Hence the same
mechanism as in QED
is found to be at work in the electroweak sector of the
standard model. In this way, we get some new insight into
symmetry aspects of a realistic theory and, concomitantly, the nature
of the observed photon.

\end{enumerate}

Clearly, the symmetry properties of combined Higgs-gauge systems are
very rich. This reflects the fact that the Higgs field contributes its own
share to the symmetries of the interacting theory.
If one works
without redundant variables, one can give a precise meaning to
breakdown of gauge symmetry: The vacuum is not invariant under ``large''
gauge transformations. Since there is no obvious order parameter
associated with such a
kind of symmetry breakdown, this concept
has never become very popular. Nevertheless, it seems to us to
have more predictive
power than the mere association of mass terms for gauge bosons with
symmetry breakdown. Thus for instance,
homotopy considerations show at once that only in abelian groups,
one can have
massless vector bosons.
In non-abelian ones, the ``large'' gauge transformations do not
have the right group structure. We immediately predict that unlike QED,
pure Yang Mills theory should have no massless vector bosons.
  From the conventional point of view, this is a mystery, since both
theories are classified as unbroken, and one has to invoke
additional mechanisms such as confinement to prevent the
appearance of massless gauge bosons.
If one finds nevertheless massless photons in non-abelian models
with simple gauge groups like the
Georgi Glashow model, a symmetry different from the standard gauge symmetry
must break
down. This is possible in Higgs models because scalar fields can increase
the number of symmetries.

The type of gauge symmetry breaking discussed in the textbooks in
connection with the Higgs mechanism cannot be directly compared
with our results, since it is discussed at a stage where the
theory still has redundant variables. We have proposed a simple
explanation why the physics consequences derived from the Higgs
mechanism in both ways are the same, in spite of conceptual differences.

\vskip 1.0cm
{\bf Acknowledgement}
\vskip 0.5cm
We would like to thank F. Lenz, S. Levit, J. Polonyi and N. Walet for
helpful discussions, and H. Grie\ss hammer and D. Lehmann for critically
reading the manuscript.
This work is supported in part by the German Federal
Minister for Research and Technology (BMFT).

\newpage
\renewcommand{\theequation}{A.\arabic{equation}}

\setcounter{equation}{0}
\section*{Appendix}
\section*{Magnetic monopoles and zeros of the adjoint Higgs field}

In this appendix, we re-derive the quantization of magnetic charge
in the Georgi Glashow model in an elementary way.
The relevant term in the magnetic field strength is the one arising from
the last term in eq. (\ref{4.11}),
\begin{equation}
{\cal U}_{\Delta}^{\dagger} \delta B_{i}^{3} \, {\cal U}_{\Delta} =
\frac{i}{g} \frac{1}{\chi^{3}} \epsilon_{ijk} \mbox{tr} \left(
\phi  [\partial_{j} \phi ,
\partial_{k} \phi ] \right) \ .
\label{A.1}
\end{equation}
Assume that the Higgs field $\phi (\vec{x}) $ vanishes at the
point $\vec{x} = \vec{x}_{0}$.
Then, in the vicinity of $\vec{x}_{0}$, we write down the Taylor expansion
\begin{equation}
\phi^{a}(\vec{x}) \simeq (\vec{x}- \vec{x}_{0})  \vec{\nabla}
\phi^{a} |_{\vec{x}=\vec{x}_{0}}
:= (\vec{x}- \vec{x}_{0}) \vec{c}^{\,a} \ .
\label{A.2}
\end{equation}
It is now easy to show that the divergence of the magnetic field develops
a $\delta$-function, indicating the presence of a magnetic monopole.
For $\vec{x} \simeq \vec{x}_{0}$, we have
\begin{equation}
{\cal U}_{\Delta}^{\dagger} \vec{\nabla}  \delta \vec{B}^{3}
{\cal U}_{\Delta}
= \frac{i}{g} \epsilon_{ijk} \partial_{i} \frac{1}{\chi^{3}}
\mbox{tr} \left( \phi [c_{j},c_{k}] \right) \ .
\label{A.3}
\end{equation}
Using
\begin{equation}
\frac{\partial}{\partial x^{i}} = c_{i}^{a} \frac{\partial}
{\partial \phi^{a}}
\label{A.4}
\end{equation}
and
\begin{equation}
\frac{\phi^{b}}{\chi^{3}} = - \frac{\partial}{\partial
\phi^{b}} \frac{1}{\chi}  \ ,
\label{A.5}
\end{equation}
we obtain
\begin{equation}
{\cal U}_{\Delta}^{\dagger} \vec{\nabla}  \delta
\vec{B}^{3} \,{\cal U}_{\Delta} =
- \frac{1}{2g} \epsilon_{ijk} \epsilon_{bcd} c_{i}^{a} c_{j}^{c} c_{k}^{d}
\frac{\partial^{2}}{\partial \phi^{a} \partial \phi^{b}} \frac{1}{\chi} \ .
\label{A.6}
\end{equation}
With the decomposition
\begin{equation}
\partial_{a} \partial_{b} = \delta_{ab} \frac{1}{3} \Delta +
( \partial_{a} \partial _{b}
- \delta_{ab} \frac{1}{3} \Delta )
\label{A.7}
\end{equation}
and
\begin{equation}
\Delta \frac{1}{r} = - 4 \pi \delta^{(3)}(\vec{r}\,)  \ ,
\label{A.8}
\end{equation}
one gets (from the scalar part of (\ref{A.7}))
\begin{equation}
{\cal U}_{\Delta}^{\dagger} \vec{\nabla}  \delta
\vec{B}^{3} \,{\cal U}_{\Delta} =
\frac{2 \pi}{3g} \epsilon_{ijk} \epsilon_{abc} c_{i}^{a}
c_{j}^{b} c_{k}^{c}
 \delta^{(3)} (\vec{\phi} ) \ .
\label{A.9}
\end{equation}
Since
\begin{equation}
\epsilon_{ijk} \epsilon_{abc} c_{i}^{a} c_{j}^{b} c_{k}^{c}
= 6 \, \mbox{det} c_{i}^{a} \ ,
\label{A.10}
\end{equation}
the prefactor in eq. (\ref{A.9}) can be used to convert the argument of the
$\delta$-function from $\vec{\phi}$ into $\vec{x}- \vec{x}_{0}$
(up to a possible sign), and
we finally obtain the correctly quantized magnetic charge,
\begin{equation}
{\cal U}_{\Delta}^{\dagger} \vec{\nabla}  \delta
\vec{B}^{3} {\cal U}_{\Delta} =
\pm \frac{4 \pi}{g} \delta^{(3)} (\vec{x}- \vec{x}_{0}) \ .
\label{A.11}
\end{equation}
The sign of the monopole strength is determined by sgn(det$c_i^a$).


\newpage
\begin{thebibliography}{99}

\bibitem{GWS}
S.L. Glashow, Nucl. Phys. {\bf 22} (1961) 579; S. Weinberg,
Phys. Rev. Lett. {\bf 19} (1967) 1264; A. Salam, Proc. 8th
Nobel Symposium Aspen\"ag\aa rden, 1968, ed. N. Svartholm
(Almqvist and Wiksell, Stockholm, 1968) p. 367.

\bibitem{Arnison}
Arnison, G. {\em et al.}, UA(1) Collaboration, Phys. Lett.
{\bf 122B} (1983) 103.

\bibitem{Englert}
F. Englert and R. Brout, Phys. Rev. Lett. {\bf 13} (1964) 321.

\bibitem{Higgs}
P. Higgs, Phys. Rev. Lett. {\bf 13} (1964) 508.

\bibitem{Guralnik1}
G. Guralnik, C. Hagen and T. Kibble, Phys. Rev. Lett.
{\bf 13} (1964) 585.

\bibitem{Kibble}
T. Kibble, Phys. Rev. {\bf 155} (1967) 1554.

\bibitem{Schwinger}
J. Schwinger, Phys. Rev. {\bf 125} (1962) 397.

\bibitem{Guralnik2}
G.S. Guralnik, Phys. Rev. Lett. {\bf 13} (1964) 295.

\bibitem{LNOT12}
F. Lenz, H.W.L. Naus, K. Ohta and M. Thies, Ann. of Phys. {\bf
233} (1994) 17, 51.

\bibitem{Georgi}
H. Georgi and S.L. Glashow, Phys. Rev. Lett. {\bf 28} (1972) 1494.

\bibitem{Kovner}
A. Kovner and B. Rosenstein, Phys. Rev.
{\bf D 49} (1994) 5571, and references therein.

\bibitem{Jackiw}
R. Jackiw, Rev. Mod. Phys. {\bf 52} (1980) 661.

\bibitem{Stoll}
D. Stoll, Phys. Lett. {\bf 336 B} (1994) 518, 524.

\bibitem{Elitzur}
S. Elitzur, Phys. Rev. {\bf D12} (1975) 3978.

\bibitem{Froehlich}
J. Fr\"ohlich, G. Morchio and F. Strocchi,
Nucl. Phys. {\bf B190} (1981) 553.

\bibitem{Dolan}
L. Dolan and R. Jackiw,
Phys. Rev. {\bf D9} (1974) 2904.

\bibitem{Lawrie}
I.D. Lawrie, Nucl. Phys. {\bf B361} (1991) 415.

\bibitem{BRST}
M. Henneaux and C. Teitelboim,``Quantization of Gauge Systems'',
Princeton University Press 1992, and references therein.

\bibitem{Hooft74}
G. 't Hooft, Nucl. Phys. {\bf B79} (1974) 276.

\bibitem{LNT}
F. Lenz, H.W.L. Naus and M. Thies, Ann. of Phys. {\bf 233}
(1994) 317.

\bibitem{Shifman}
F. Lenz, M. Shifman and M. Thies, MIT preprint CTP-2391, Dec. 1994,
hep-th/9412113.

\bibitem{Coleman}
S. Coleman, in: ``Aspects of Symmetry'', Cambridge 1985, ch. 5, p. 121.

\bibitem{Polyakov}
A.M. Polyakov, JETP Lett. {\bf 20} (1974) 194.

\bibitem{Rossi}
P. Rossi, Phys. Rep. {\bf 86} (1982) 317.

\bibitem{Arafune}
J. Arafune, P.G.O. Freund and C.J. Goebel, J. Math. Phys. {\bf 16}
(1975) 433.

\bibitem{Witten}
E. Witten, Phys. Lett. {\bf 86B} (1979) 283.

\bibitem{Strocchi}
F. Strocchi, ``Elements of quantum mechanics of infinite systems'',
Int. School for Advanced Studies Lecture Series No. 3, World Scientific
(Singapore 1985).

\bibitem{Fradkin}
E. Fradkin and S.H. Shenker, Phys. Rev. {\bf D19} (1979) 3682.

\bibitem{Banks}
T. Banks and E. Rabinovici, Nucl. Phys. {\bf B160} (1979) 349.

\bibitem{Dimopoulos}
S. Dimopoulos, S. Raby and L. Susskind, Nucl. Phys. {\bf B173} (1980)
208.

\bibitem{Ovrut}
B.A. Ovrut, J. Math. Phys. {\bf 19} (1978) 418.

\bibitem{Baal}
P. van Baal, ``Gauge theory in a finite volume'', Lecture
presented at the XXVIII Cracow School of Theoretical Physics,
May 31 - June 10, 1988.

\bibitem{Kogut}
J. Kogut and L. Susskind, Phys. Rev. {\bf D11} (1975) 395.

\bibitem{Leutwyler}
H. Leutwyler, ``Goldstone Bosons'', Bern preprint BUTP-94/17.

\bibitem{Montanet}
Review of particle properties, L. Montanet {\em et al.},
Phys. Rev. {\bf D50} (1994) 1179.

\end{thebibliography}
\end{document}